\documentclass{article}
  
\usepackage[utf8]{inputenc}
\usepackage[T1]{fontenc}
\usepackage[mathscr]{euscript}
\usepackage{geometry}
\usepackage{amssymb}
\usepackage{amsmath}
\title{General Relativity coupled with Non-Linear Electrodynamics: results and limitations}
\author{Stefano Chinaglia\footnote{e-mail: s.chinaglia@unitn.it} \\
\smallskip \\
\it Dipartimento di Fisica, Università di Trento \\
\it Via Sommarive 14, 38123 Trento, Italia \\
\smallskip \\
\it TIFPA (INFN) \\
\it Via Sommarive 14, 38123 Trento, Italia}
\date{\today}

\begin{document}
\maketitle
\begin{abstract}

We discuss how to generate a black hole solution of the Einstein Equations (EE) via non-linear electrodynamics (NED). We discuss the thermodynamical properties of a general NED solution, recovering the First Law. Then we illustrate the general mechanism and discuss some specific cases, showing that finding a generating Lagrangian (for a specific solution) only requires solving an algebraic equation (we study some analytical cases). Finally, we argue that NED paradigm, though self-consistent, is not the best tool for studying regular black holes.

\end{abstract}

\section{Introduction}

Schwarzschild solution is the very first solution of the Einstein Equations and the most paradigmatic metric for a black hole. However Schwarzschild metric has a problem: when $r \rightarrow 0$ it blows up into a singularity. Since one expects nature to be singularity-free (a Born-Infeld inspired approach \cite{Born_Infeld}), one should conclude that Schwarzschild solution has to be corrected near the origin, in order to avoid the singularity.

The first author to propose a regular solution was Bardeen, fifty years ago \cite{Bardeen}. Then many other came, spanning along the decades: just to say a few \cite{Bronnikov} \cite{Elizalde} \cite{Dymnikova} \cite{Hayward_1} \cite{NSS} \cite{ANSS} \cite{Modesto} \cite{CMSZ} \cite{Dymnikova_2} \cite{Culetu} \cite{Maeda} \cite{Horava} \cite{Horava_1} \cite{KS} \cite{CCO} \cite{Pradhan} \cite{Ma} \cite{Johannsen} \cite{Rodrigues}. Among the huge variety of adopted techniques, two main paths have been followed, in order to find regular solutions: the first modifies the geometric sector of the Einstein Equations, the second works on the matter content, looking for some (typically exotic) matter, able to regularize the central singularity.

Among these, \cite{Elizalde} \cite{NSS} \cite{ANSS} \cite{Modesto} \cite{CMSZ} \cite{Maeda} \cite{Horava} \cite{Horava_1} \cite{KS} \cite{CCO} work on the gravitational sector. \cite{NSS} \cite{ANSS} smear the singularity introducing non commuting coordinates, i.e. $[x_{\mu}, x_{\nu}] \neq \delta_{\mu \nu}$, while \cite{CMSZ} \cite{Maeda} \cite{Horava} \cite{Horava_1} \cite{KS} \cite{CCO} directly attack the Einstein tensor. In particular, \cite{CCO}  presents the solution

\begin{equation}
\label{CCO}
f(r) = 1 - \frac{r^2}{4\xi} \left(1 \pm \sqrt{1 + \frac{16\xi^2}{r^4} \left( 1 + \frac{c_1}{\xi} + c_2 r \right)} \right)
\end{equation}

together with its dynamical origin (we remand to it for details). Unfortunately, solution (\ref{CCO}) is still singular. On the other hand, \cite{Bronnikov} \cite{Dymnikova} \cite{Dymnikova_2} \cite{Culetu} \cite{Pradhan} follow the second approach, working on NED.

In the late '90 E. Ayon-Beato and A. Garcia proposed a scheme to generate solutions (even regular solutions) from a non linear electrodynamical Lagrangian minimally coupled to the standard Lagrangian of General Relativity (GR) \cite{Beato}. They successfully applied their scheme to build up a modified Reissner-Nordström, Bardeen-like, solution, later identified as a magnetic monopole solution for the Einstein Equations \cite{Beato_1}. The scheme of Beato and Garcia was later generalized by I. Dymnikova \cite{Dymnikova} for spherical and static solutions.

Unfortunately, that scheme is very difficult to be implemented in practice and very rarely produces full analytical solutions (we mean, regular solution + explicit generating Lagrangian). This paper has three main aims:

\begin{enumerate}
\item{simplify the general problem, producing a relatively simple way to find the generating Lagrangian (given the solution);}
\item{provide some examples, for particularly simple (but non trivial) cases;}
\item{show that the NED model provides a class of regular solutions, but the most general solution is still singular.}
\end{enumerate}

Finally we should mention an "intermediate" way, which leaves untouched the Einstein tensor, but introduces a non minimal coupling between the gravitational and the matter Lagrangian: on this way are the works by Dereli and Sert \cite{Dereli}, Sert \cite{Sert}, Balakin and Lemos \cite{Balakin}, Balakin and Zayats \cite{Zayats} and, recently, Balakin, Lemos and Zayats \cite{BLZ}. \cite{Dereli} and \cite{Sert} write the Lagrangian in the form $\mathscr{L} = R - Y(R)I$, where $Y(R)$ is a given function of $R$, while \cite{Balakin}, \cite{Zayats} and \cite{BLZ} follow the lines of Horndeski \cite{Horndeski} and of Drummond and Hathrell \cite{Drummond} and use $\mathscr{L} = R + F^{(a)}_{\mu\nu} F^{(a)\mu\nu} + \mathscr{R}^{\mu\nu\rho\sigma} F^{(a)}_{\mu\nu} F^{(a)}_{\rho\sigma}$, where $\mathscr{R}^{\mu\nu\rho\sigma}$ is a complicated object worked out from the Riemann tensor. While the approach of \cite{Dereli} and \cite{Sert} is not able to produce regular black holes (as we will discuss), that of \cite{BLZ} actually does.

The paper is organised as follows: in sec. 2 we review the scheme of Beato-Garcia-Dymnikova (BGD), while in sec. 3 we show that, under certain conditions, it produces unique solutions; in sec. 4 we recover the First Law of thermodynamics and in sec. 5 we provide some explicit solutions; in sec. 6 we show the limitations of BGD procedure and finally we discuss our results in sec. 7. The two appendices contain technical details: app. A shows that a specific choice for the Lagrangian implies a single choice for the Hamiltonian and vice versa; app. B calculates the quasi local energy function $E_{ADM}(r)$.

\section{Framework and equations of motion}

We start considering the approach by E. Ayon-Beato and A. Garcia \cite{Beato} and generalizing it to any radially symmetric metric of the form

\begin{equation}
\label{metric}
ds^2 = -f(r,t)dt^2 + \frac{1}{f(r,t)} dr^2 + r^2 d\Omega^2_k
\end{equation}

where $f(r,t)$ is some arbitrary function of $r$ and $t$ and $d\Omega^2_k$ is the volume of a 2-sphere ($k=1$) or a hyperboloid ($k=-1$) or a torus ($k=0$). Although topology doesn't play any role in our argument, topological black holes have been widely studied since the works of Brill, Luoko and Peldàn \cite{BLP}, Mann \cite{Mann} and Vanzo \cite{Vanzo} and there is no effort to generalize the discussion of the spherical case to different topologies. The black hole horizon $r_H$ is given by the condition $f(r_H,t)=0$. The general formalism of a static and spherical NED solution has been already discussed by I. Dymnikova in \cite{Dymnikova}.

In the NED picture, a specific solution of the EE is generated by an action

\begin{equation}
\label{action}
\mathscr{I} = \frac{1}{16\pi} \int d^4 x \ \sqrt{-g} (R - 2\Lambda - 4 \mathscr{L}(I))
\end{equation}

where $R$ is the Ricci scalar, $\Lambda$ is a cosmological constant $I=\frac{1}{4}F^{\mu\nu} F_{\mu\nu}$ is an electromagnetic-like tensor and $\mathscr{L}(I)$ is a suitable function (whose choice depends on the specific choice of $f$). As in standard electrodynamics, tensor $F_{\mu\nu}$ is written antisymmetrizing some vector potential: $F_{\mu\nu}=\partial_\mu A_\nu - \partial_\nu A_\mu$; however, since $A_\mu$ never appears alone into our calculation, we will simply take $F_{\mu\nu}$ and $I$ as "true" dynamical variables. We also define the new objects $P_{\mu\nu}\equiv F_{\mu\nu} (\partial_I \mathscr{L}(I))$, $P \equiv \frac{1}{4} P_{\mu\nu}P^{\mu\nu}$ and $\mathscr{H} \equiv 2I(\partial_I \mathscr{L}(I)) - \mathscr{L}(I)$.

Now, action (\ref{action}) generates the following equations of motion:

\begin{equation}
\label{eq_motion_1}
G^\nu_\mu + \Lambda \delta^\nu_\mu = 4g^{\alpha\nu} \frac{\partial \mathscr{L}}{\partial g^{\mu\alpha}} - 2 \mathscr{L}\delta^\nu_\mu 
\end{equation}

\begin{equation}
\label{eq_motion_2}
\nabla^\mu (F_{\mu\nu} \partial_I \mathscr{L}) = 0
\end{equation}

which read, switching variables,

\begin{equation}
\label{eq_motion_1P}
G^\nu_\mu + \Lambda \delta^\nu_\mu = 2((\partial_P \mathscr{H})P_{\mu\rho}P^{\nu\rho} + (\mathscr{H} - 2(\partial_P \mathscr{H})P) \delta_\mu^\nu)
\end{equation}

\begin{equation}
\label{eq_motion_2P}
\nabla^\mu P_{\mu\nu} = 0
\end{equation}

In both cases, the first is the usual set of EE, while the other is the set of the (nonlinear) Maxwell Equations (ME). The key trick is to solve first the ME, then plug their result into the EE. Perform so the following ansatz: $P_{\mu\nu} = (\delta_\mu^0 \delta_\nu^1 - \delta_\nu^0 \delta_\mu^0) D(r,t)$, where $D(r,t)$ is some function to be determined. This makes the ME become

\begin{equation}
\label{D_1}
\frac{\partial}{\partial r} D(r,t) + \frac{2}{r} D(r,t) = 0
\end{equation}

\begin{equation}
\label{D_2}
\frac{\partial}{\partial t} D(r,t) = 0
\end{equation}

for $\nu=0,1$ respectively; $\nu=2,3$ just produce trivial identities.

The second equation simply states that the approach of \cite{Beato} only generates static solutions: i.e. the scheme of \cite{Dymnikova} is already the most general spherical case. They give $D(r)=K/r^2$, where $K$ is an integration constant. Notice that both eq. (\ref{D_1}) and (\ref{D_2}) do not keep trace of the choice of the metric: this is somewhat expected, since the substitution $I \rightarrow P$ kills the dependence from $\mathscr{L}$ into the ME (in any case, the choice of the metric is still kept by the constant $K$). Eq. (\ref{D_1}) and (\ref{D_2}) then tell us that

\begin{equation}
\label{P(r)}
P(r) = -\frac{1}{4} D^2 (r) = - \frac{K^2}{2r^4}
\end{equation}

and, most of all, they tell (inverting $P(r)$) how to find $r(P)$.
Now that we solved the ME, we pass to the EE. The only non trivial component is that of $G_t^t$:

\begin{equation}
\label{Gtt}
G^t_t + \Lambda = 2 \mathscr{H} \ \ \ \ \ \ \ \ \ \ \Rightarrow \ \ \ \ \ \ \ \ \ \ \mathscr{H} = \frac{f'r + f - k + \Lambda r^2}{2r^2}
\end{equation}

(Note that $\mathscr{H}$ in eq. (\ref{Gtt}) is viewed as a function of $r$: $\mathscr{H}=\mathscr{H}(P)=\mathscr{H}(P(r))$; other changes in notation are obvious.)

Eq. (\ref{Gtt}) is a linear, inhomogeneous, differential equation. Assuming $\mathscr{H}$ to be given, it is solved using simple theory of linear, inhomogeneous, differential equations:

\begin{equation}
\label{metric_solution}
f(r) = k - \frac{2c}{r} - \frac{\Lambda}{3} r^2 + \frac{2}{r} \int r^2 \mathscr{H}(r) \ dr
\end{equation}

where $c$ is an integration constant, which will be set so that $f(r \rightarrow \infty) = k - \frac{2m}{r} - \frac{\Lambda}{3} r^2 + o(r^{-2})$, with $m$ the physical mass of the black hole (take care of it, because it will be useful later). We can also write eq. (\ref{metric_solution}) in a more standard shape:

\begin{equation}
\label{f_M}
f(r) = k - \frac{2M(r)}{r} - \frac{\Lambda}{3} r^2
\end{equation}

where $M(r) \equiv c - \int r^2 \mathscr{H}(r) \ dr$.

(By the way, eq. (\ref{metric_solution}) limits the behaviour of $\mathscr{H}(r)$: indeed, in order to recover the classical result at infinity, we must require $\mathscr{H} \ll 1/r^3$ at infinity.)

Unfortunately, eq. (\ref{metric_solution}) requires the Hamiltonian to be known, not the Lagrangian. The relation between the Lagrangian and Hamiltonian, though unique, is in general quite complicated. Both finding the solution, given the Lagrangian, or the generating Lagrangian, given the solution lead to very difficult differential equations.

Before concluding this section, let us check if the Weak Energy Condition (WEC) is fulfilled. However the answer is, in general, inconclusive: indeed, the WEC is satisfied if and only if \cite{Hawking_Ellis}

\begin{align}
\label{WEC}
& \rho \geq 0 \\
& \rho + p_k \geq 0 \ \ \ \ \ k=1,2,3
\end{align}

where $\rho = T_t^t$ is the energy density and $p_k = - T_k^k$ the principal pressures. They result in three conditions over our Hamiltonian:

\begin{align}
\label{WEC_1}
& 2 \mathscr{H} \geq 0 \\
\label{WEC_2}
& 2 \mathscr{H} - 2 \mathscr{H} \geq 0 \\
\label{WEC_3}
& - \mathscr{H} r \geq 0
\end{align}

As one can see, the second condition is trivially satisfied. The problem is that we have no way to answer if the remaining two are satisfied or not: indeed the Hamiltonian contains (at least) one completely free parameter, so that we are not able to tell in general the overall sign of $\mathscr{H}(r)$ and $\mathscr{H}'(r)$.

\section{Uniqueness of the solution}

One may wonder if the procedure described above produces a unique solution or if more solutions are available for the same Lagrangian: in principle this is possible, due to the non trivial relation between $\mathscr{H}$ and $\mathscr{L}$. We discuss in Appendix A that fixing the Lagrangian fixes the Hamiltonian and vice versa, so here we just need to show that, keeping fixed the Lagrangian or the Hamiltonian, the generated solution (\ref{metric_solution}) is unique. For calculative convenience, we prefer to work with the Hamiltonian, so have the following assumptions:

\begin{enumerate}
\item{the Hamiltonian $\mathscr{H}(P)$ is kept fixed;}
\item{any metric has the same asymptotic behaviour: i.e. $f(r) \rightarrow k-\frac{2m}{r}-\frac{\Lambda}{3}r^2$;}
\item{any solution has the same $\Lambda$ and the same ADM mass.}
\end{enumerate}

Suppose $\mathscr{H}$ generating two solutions, $f_1$ and $f_2$. Since the Hamiltonian is the same and both must obey to eq. (\ref{Gtt}), they differ at most for a Schwarzschild term:

\begin{equation}
\label{Delta_f}
\Delta f' = - \frac{\Delta f}{r} \ \ \ \ \ \ \ \ \ \ \Rightarrow \ \ \ \ \ \ \ \ \ \ \Delta f = \frac{A}{r}
\end{equation}

where $\Delta f = f_2 - f_1$ and $A$ is some integration constant. Applying now assumption 2, we write both $f_i$ in the form $f_i (r) = k - \left( \frac{2m}{r} + \frac{\Lambda}{3} r^2 \right) g_i (r)$, where the $g_i (r)$ have to be determined, but with the requirement of $g_i (r) \rightarrow 1$ for $r \rightarrow \infty$. From eq. (\ref{Delta_f}) immediately turns out that, if the space is asymptotically flat (i.e. $\Lambda=0$), then $f_1=f_2$: indeed, making use of assumptions 2 and 3, one has

\begin{equation}
\begin{split}
\label{uniqueness_1}
f_2(r) = f_1(r) + \frac{A}{r} \ \ \ \ \ \ \ \ \ \ &\Rightarrow \ \ \ \ \ \ \ \ \ \ k - \frac{2m}{r} g_2(r) - \frac{\Lambda}{3} r^2 g_2(r) = k - \frac{2m}{r} \left( g_1(r) - \frac{A}{2m} \right) - \frac{\Lambda}{3} r^2 g_2(r) \\
&\Rightarrow \ \ \ \ \ \ \ \ \ \ g_2(r) - g_1(r) = - \frac{A}{2m \left( 1 + \frac{\Lambda}{6m} r^3 \right)} \\
\end{split}
\end{equation}

If $\Lambda \neq 0$, this difference fulfills the condition that $g_1$ and $g_2$ have the same limit for any $A$. On the other hand, if the space is asymptotically flat, the only way to fulfill the same condition is to ask $A=0$. So, only in the case of asymptotic Minkowski space, NED generates unique solutions.

\section{Thermodynamical properties: the First Law}

Here we study the basic thermodynamical properties of a general NED black hole. Detailed studies have been carried out for some specific black holes: among the others \cite{Pradhan} \cite{BLP} \cite{Gibbons} \cite{Park} \cite{Akbar} \cite{Saneesh}. Since action (\ref{action}) is written as the integral of $R+something$ (which does not depend on $R$), the entropy of the resulting black hole is just proportional to its area: using Hawking temperature \cite{Hawking} as temperature $T$ of the system, a metric of the form (\ref{metric}) gives

\begin{equation}
\label{first_law_1}
TdS = \frac{1}{2} f'(r_H) r_H dr_H
\end{equation}

where $S$ is the entropy. In order to identify the rhs of this equation with some thermodynamical quantity, we make use of the Hayward method \cite{Hayward}, evaluating the equation of motion on the horizon. Go back to eq. (\ref{eq_motion_1}) and write the only non trivial component of the EE:

\begin{equation}
\begin{split}
\label{first_law_2}
G_0^0 + \Lambda = 4g^{00} \frac{\partial \mathscr{L}}{\partial g^{00}} - 2\mathscr{L} \ \ \ \ \ \ \ \ \ \ &
\Rightarrow \ \ \ \ \ \ \ \ \ \ \frac{f'r + f - k}{r^2} + \Lambda = -2 \left(\frac{2f}{f'} \frac{\partial \mathscr{L}}{\partial r} + \mathscr{L} \right) \\
& \Rightarrow \ \ \ \ \ \ \ \ \ \ f'(r_H) = \frac{k - \Lambda r^2_H - 2r_H^2 \tilde{\mathscr{L}}(r_H)}{r_H} \\
& \Rightarrow \ \ \ \ \ \ \ \ \ \ \frac{1}{2} f'(r_H) r_H dr_H = \frac{1}{2} \left( k - \Lambda r^2_H + \frac{1}{2} (f''(r_H)r_H + 2f'(r_H)) \right) dr_H
\end{split}
\end{equation}

We recall that $g^{00} \partial_{g^{00}} = 1/f \partial_{1/f} = - f/f' \partial_r$ and we also notice that

\begin{equation}
\label{limit}
\begin{split}
\frac{f(r_H)}{f'(r_H)} &= \lim_{r \rightarrow r_H} \frac{f(r)}{f'(r)} \\
&= \lim_{r \rightarrow r_H} (r-r_H) \frac{\frac{1}{2} f''(r_H) (r-r_H) + \frac{1}{6} f'''(r_H) (r-r_H)^2 + ...}{f''(r_H) (r-r_H) + \frac{1}{2} f'''(r_H) (r-r_H)^2 + ...} \\
&=0
\end{split}
\end{equation}

whatever is the ratio of the two polynomials (the same argument holds also when other derivatives of $f$ vanish on the horizon). This allows us to write the last line of eq. (\ref{first_law_2}).

Now, the lhs of this equation is clearly $TdS$, while the rhs, at a first look, seems to be the sum of the (differential) black hole internal energy and matter energy contribution, $dU_{bh}+dU_{\Lambda}+dU_{matter}$: one recognises the Misner-Sharp energy $E_{MS}=\frac{k}{2}r_H$ plus some matter contribution (indeed, if we go back in vacuo and set to 0 the matter Lagrangian, we just recover the standard internal energy $U_{bh}=km$).

In order to give a robust foundation to this suggestion, let us recall the expression for the ADM quasi local energy function $E(r)=\frac{1}{2} r^2 f'(r)$ found in Appendix B. Differentiating $E(r)$ and evaluating it on the horizon, one gets

\begin{equation}
\label{ADM_energy}
E'(r_H)dr_H = \frac{1}{2} (f''r_H^2 + 2f'r_H) dr_H
\end{equation}

Comparing it with eq. (\ref{first_law_2}), it is now clear that $U_{matter}=\frac{1}{2}(E(r_H )+constant)$, where the constant is tuned so that $U_{matter}=0$ in vacuo (or in presence of $\Lambda$ alone). This is obtained by simply subtracting the ADM mass; so that, if

\begin{equation}
\label{internal_energy}
U = U_{bh} + U_{\Lambda} + U_{matter} \ \ \ \ \ \ \ \ \ \ \Rightarrow \ \ \ \ \ \ \ \ \ \ U = E_{MS} - \frac{\Lambda}{6} r_H^3 + \frac{1}{2} (E(r_H) - E_{ADM})
\end{equation}

Then eq. (\ref{first_law_2}) exactly resembles the form of the first law, i.e. $TdS=dU$. This holds for any metric
in the form (\ref{metric}) obeying to action (\ref{action}).

\section{Exact solutions}

It is very difficult to get an exact solution of a NED model (if we exclude some trivial solutions, as Schwarzschild (KS) or Reissner-Nordström (RN)). And it's even more difficult finding a \it regular \rm solution. In this section we present a scheme which allows to build up (even regular) solutions from a given Lagrangian (or vice versa), then we will focus on the only solvable cases.

\subsection[toctitle]{General algorithm for the solution}

We need how to write $\tilde{\mathscr{L}}(r) \equiv (\mathscr{L} \circ I)(r)$ in terms of the solution. The first step is to write $\partial_I \mathscr{L}$, $I$ and $\partial_I$ as functions of $r$ alone.  Making use of the relations between $P$ and $r$ and between $\partial_I \mathscr{L}$ and $\partial_I \mathscr{H}$, we have that

\begin{equation}
\label{rel_1}
\frac{\partial}{\partial P} = \frac{r^5}{8K^2} \frac{\partial}{\partial r} \ \ \ \ \ \ \ \ \ \ \Rightarrow \ \ \ \ \ \ \ \ \ \ \partial_I \mathscr{L} = \frac{16K^2}{r^2 \left( f''r^2 -2(f-k) \right)}
\end{equation}

Then, using this and the relation between $P$ and $I$, we have that

\begin{equation}
\label{rel_2}
\begin{split}
P = I (\partial_I \mathscr{L})^2 \ \ \ \ \ &\Rightarrow \ \ \ \ \ I = - \frac{\left( f''r^2 - 2(f-k) \right)^2}{128K^2} \\
&\Rightarrow \ \ \ \ \ \frac{\partial}{\partial I} = - \frac{64K^2}{\left( f''r^2 - 2(f-k) \right)(f'''r^2 + 2f''r - 2f')} \frac{\partial}{\partial r}
\end{split}
\end{equation}

Finally, collecting things together (i.e. plugging eq. (\ref{rel_2}) into eq. (\ref{rel_1})) we find how to write the Lagrangian as a function of r:

\begin{equation}
\label{L(r)}
\begin{split}
\tilde{\mathscr{L}}(r) & = \int \partial_{I(r)} (\mathscr{L} \circ I)(r) \ dI(r) \\
& = - \frac{f''r + 2f' + 2\Lambda r}{4r}
\end{split}
\end{equation}

Notice the $\Lambda$-term at the numerator; it just comes out from the integration constant, tuned so that, for $\tilde{\mathscr{L}}(r)=0$, the solution reduces to a pure SdS (same form of eq. (\ref{f_M}), but with $M(r)=constant$).

With this we are done, as we only need to assume $\tilde{\mathscr{L}}(r)$ to be known: if so, the problem of finding $f(r)$ is reduced to solve an ordinary second order inhomogeneous differential equation\footnote{Notice, however, that in eq. (\ref{L(r)}) only the first and second derivative of $f$ appear - not $f$ itself. In practice this means solving just a first order equation + an integral.}:

\begin{equation}
\label{eq_f(r)}
f''(r)r + 2f'(r) + 2\Lambda r + 4r \tilde{\mathscr{L}}(r) = 0
\end{equation}

The resulting function is exactly the same that found in eq. (\ref{metric_solution}) and the procedure is completely equivalent. In alternative, one can just use eq. (\ref{metric_solution}) to fix the $f(r)$ and use then eq. (\ref{eq_f(r)}) to find the associated $\tilde{\mathscr{L}}(r)$; the scheme can be walked in both directions.

The only thing we are left with, in order to write explicitly also the generating Lagrangian, is to find $r(I)$, then put it into $\tilde{\mathscr{L}}(r)$. In order to do this, recall the definition of $P(I)$ and its expression in terms of $r$ (eq. (\ref{P(r)})). It is easy to see that

\begin{equation}
\label{I(r)}
\frac{\partial \sqrt{-8K^2 I}}{\partial r} = r^2 (\partial_r \tilde{\mathscr{L}}(r)) \ \ \ \ \ \ \ \ \ \ \Rightarrow \ \ \ \ \ \ \ \ \ \ \sqrt{-8K^2 I} = \int r^2 (\partial_r \tilde{\mathscr{L}}(r)) \ dr
\end{equation}

but, since $\tilde{\mathscr{L}}(r))$ is known, the last integral can be performed. The only thing to do is to find a good choice of
$\tilde{\mathscr{L}}(r))$, so that integral (\ref{I(r)}) can be solved and, most of all, the resulting function can be inverted.

In order to do this, it is clear that (unless some very lucky cases), $\tilde{\mathscr{L}}(r))$ should be written in a polynomial
shape: if so, eq. (\ref{I(r)}) is an algebraic equation and there's a hope to solve it (even explicitly, if it is of degree less or equal than 4).
This is not strictly necessary, but if not, either the resulting equation becomes transcendental or $\int \partial_r \tilde{\mathscr{L}}(r)$
becomes very difficult to be performed.

\subsection[toctitle]{Example 0}

Here we show that the topological Schwarzschild-de Sitter-Reissner-Nordström (SdSRN) solution, namely

\begin{equation}
\label{SdSRN_solution}
f(r) = k - \frac{2m}{r} + \frac{q^2}{r^2} - \frac{\Lambda}{3} r^2
\end{equation}

is actually generated by the general algorithm. This is a must, since any modification of the electromagnetism is expected to reduce
asymptotically to Maxwell theory. So, assume $\mathscr{L}(I)=I$: if so, one gets $P=I$ and $\mathscr{H}(P)=\mathscr{L}(I)$: in this case, it is easy then to see that eq. (\ref{Gtt}) and eq. (\ref{eq_f(r)}) are equivalent: indeed, derivating eq. (\ref{Gtt}) one gets

\begin{equation}
\begin{split}
\label{H=L}
f''r + 2f' + 2\Lambda r = 4r \mathscr{H} +2r^2 \partial_r \mathscr{H}(P(r)) \ \ \ \ \ \ \ \ \ \ &\Rightarrow \ \ \ \ \ \ \ \ \ \ 
f''r + 2f' + 2\Lambda r = 4r \left( -\frac{K^2}{2r^4} \right) +2r^2 \left( \frac{4K^2}{2r^5} \right) \\
&\Rightarrow \ \ \ \ \ \ \ \ \ \ f''r + 2f' + 2\Lambda r = - 4r \mathscr{L}(I(r))
\end{split}
\end{equation}

Shown this, we only need to prove that just one of the eq. (\ref{Gtt}) or (\ref{eq_f(r)}) provide the SdSRN solution. Work on eq. (\ref{Gtt}): using our result for $P(r)$ we have

\begin{equation}
\begin{split}
\label{SdSRN_equation}
f(r) &= k - \frac{2c}{r} - \frac{\Lambda}{3} r^2 + \frac{2}{r} \int dr \ r^2 \left( -\frac{K^2}{2r^4} \right) \\
&= k - \frac{2c}{r} - \frac{\Lambda}{3} r^2 + \frac{K^2}{r^2}
\end{split}
\end{equation}

With the obvious identifications $c=m$ and $K=q$ this is exactly the SdSRN solution. The classical result (although still singular)
has been recovered.

\subsection[toctitle]{Example 1}

Consider the following Lagrangian:

\begin{equation}
\label{Lagrangian_1}
\tilde{\mathscr{L}}(r) = \frac{A}{(\xi^2 + r^2)^2}
\end{equation}

where $A$ is some free parameter and $\xi$ is a parameter with dimension of a length. Eq. (\ref{I(r)}) tells then that

\begin{equation}
\label{algebraic_equation_example_1}
\sqrt{-8K^2 I} = A \frac{(\xi^2 + 2r^2)}{(\xi^2 + r^2)^2}
\end{equation}

This is an algebraic equation of second degree in $r^2$. After some boring but straightforward managing, we get

\begin{equation}
\label{r(I)_example_1}
r(I) = \left(\frac{1 - \frac{\xi^2}{A} \sqrt{-8K^2 I}}{-8K^2 I} \right)^{1/4} \sqrt{A} \sqrt{1 + \sqrt{1 - \frac{\xi^2}{A} \sqrt{-8K^2 I}}}
\end{equation}

Notice the implicit "+" in front of the first term in the square root: since eq. (\ref{algebraic_equation_example_1}) is quadratic,
one would expect both $\pm$ solutions. However we are forced to choose the "+" and discard the "$-$", since otherwise
$r \not\in \mathbb{R}$. Inserting this result in eq. (\ref{Lagrangian_2}) one gets

\begin{equation}
\label{Lagrangian_1_I}
\mathscr{L}(I) = - \frac{8 K^2}{A} \frac{I}{\left(1 + \sqrt{1 - \frac{\xi^2}{A} \sqrt{-8K^2 I}} \right)^2}
\end{equation}

The only thing we are left with is to find the solution associated to Lagrangian (\ref{Lagrangian_1_I}). Using eq.
(\ref{Lagrangian_1}) and eq. (\ref{L(r)}) one solves the differential equation and finds

\begin{equation}
\label{f_example_1}
f(r) = a - \frac{b}{r} -\frac{\Lambda}{3} r^2 + \frac{2A}{\xi} \frac{\arctan{\left(\frac{r}{\xi}\right)}}{r}
\end{equation}

where $a$ and $b$ are suitable integration constants. (Notice that this result is the same, unless different names for the parameters,
to that found in \cite{Rinaldi}. Their approach was completely different than ours, so it may be just a mere coincidence, however this
fact deserves a further investigation.)
We set $a=k$ and, using the mass fixing condition, we find $m = b/2+\pi A/2\xi$. Of course it doesn't fix the parameter $b$ (as we will discuss later, this is a weak point of the BGD approach).

Finally, note that this would be a singular solution also if we could fix $b=0$: in that case, indeed, $f(r)$ would be (near the origin)

\begin{equation}
\label{f_example_1_origin}
f(r) = k + \frac{2A}{\xi^2} - \left( \frac{\Lambda}{3} + \frac{2A}{3\xi^3} \right) r^2
\end{equation}

but Sakharov criterion \cite{Sakharov} states that a black hole has a regular centre if $f(r \rightarrow 0)$ goes at least as $f \sim 1 - a r^2$. So eq. (\ref{f_example_1}) in any case does not represent a regular black hole. Notice also that, since $b$ cannot be fixed, even $A$ is completely free, so (recalling conditions (\ref{WEC_1})-(\ref{WEC_3})) we cannot say if the WEC is (in general) satisfied or not.

\subsection[toctitle]{Example 2}

Consider the following Lagrangian:

\begin{equation}
\label{Lagrangian_2}
\tilde{\mathscr{L}}(r) = A \frac{(\xi^2 - r^2)}{(\xi^2 + r^2)^3}
\end{equation}

where everything has the same meaning of eq. (\ref{Lagrangian_2}). Eq. (\ref{I(r)}) this times tells that

\begin{equation}
\label{algebraic_equation_example_2}
\sqrt{-8K^2 I} = \frac{8Ar^4}{(\xi^2 + r^2)^3}
\end{equation}

This is an algebraic equation of third degree in $r^2$. Also this equation can be solved exactly and its solution put into eq. (\ref{Lagrangian_2})
in order to find $\mathscr{L}(I)$ (we don't write it explicitly here, since it is an unnecessary, while long and complicated expression). The only thing we are left with is to find the solution $f(r)$: using the same procedure applied in the previous subsection, we find

\begin{equation}
\label{f_example_2}
f(r) = a - \frac{b}{r} - \frac{\Lambda}{3} r^2 + \frac{4A}{\xi^2 + r^2} + \frac{4A}{\xi} \frac{\arctan{\left(\frac{r}{\xi}\right)}}{r}
\end{equation}

In order to get a SdS solution, we should ask $a=k$ and $m = b/2 + 2\pi A/\xi$; once again, parameter $b$ hasn't been fixed, but once again regularity requires $b=0$; the $\Lambda = 0$, $k=1$, $b=0$ case (with the appropriate choice for $A$) has been already found by I. Dymnikova \cite{Dymnikova}.

Finally, in order to satisfy the WEC, applying conditions (\ref{WEC_1})-(\ref{WEC_3}), one easily sees that $A \leq 0$ is needed. However, when we look for the asymptotic form of (\ref{f_example_2}), we have

\begin{equation}
\label{WEC_ex_2}
f(r \rightarrow \infty) = k - \frac{2 \left( \frac{b}{2} + \frac{\pi A}{\xi} \right)}{r} - \frac{\Lambda}{3} r^2 - \frac{4A}{r^2} + o(r^{-3})
\end{equation}

so $\sqrt{A}$ plays the role of charge $Q$, requiring then $A \geq 0$. As found in \cite{Dymnikova} (where the study is more detailed), solution (\ref{f_example_2}) actually satisfies the WEC.

\subsection[toctitle]{Example 3}

Consider the following Lagrangian:

\begin{equation}
\label{Lagrangian_3}
\tilde{\mathscr{L}}(r) = A \frac{r^4 - 7\xi^2 r^2 -5 \xi^4}{(\xi^2 + r^2)^4}
\end{equation}

Eq. (\ref{I(r)}) this times tells that

\begin{equation}
\label{algebraic_equation_example_3}
\sqrt{-8K^2 I} = A \frac{5\xi^6 + 20\xi^4 r^2 + 17\xi^2 r^4 - 4r^6}{2(\xi^2 + r^2)^4}
\end{equation}

This time we have an algebraic equation of fourth degree in $r^2$. The equation can still be solved explicitly, but only at a price of a very intricate result (which we don't write here). We only calculate the solution generated by this Lagrangian: once again, we find

\begin{equation}
\label{f_example_3}
f(r) = a - \frac{b}{r} - \frac{\Lambda}{3} r^2 + \frac{2A\xi^2}{(\xi^2 + r^2)^2} + \frac{2A}{\xi} \frac{\arctan{\left(\frac{r}{\xi}\right)}}{r}
\end{equation}

with $a=k$ and $m = b/2 + \pi A/2\xi$: once again, $b$ is not fixed. Unlike solution (\ref{f_example_2}), this is a new solution, generated by an exact NED Lagrangian\footnote{We recall that we don't write the explicit $\mathscr{L}(I)$ just for a matter of convenience. However an exact Lagrangian is available.} $\mathscr{L}(I)$. Finally, as in example 1, we cannot say if the WEC is satisfied or not, since $A$ can take any value (depending on the choice of $b$, since we must fulfill the condition $b - \pi A / \xi \geq 0$, i.e. the mass of the black hole is positive).

\section{Limitations of the BGD procedure}

In a quantum theory of gravity, one would expect divergences to be smoothed out by quantum effects. A similar behaviour is expected to hold
also for semiclassical models, where suitable modifications are added to the classical theory. In this section we show that the BGD (but also
other similar procedures) is not able to achieve this goal in general.

Consider a metric in the form (\ref{metric}), generated by an action

\begin{equation}
\label{action_general}
\mathscr{I} = \frac{1}{16\pi} \int d^4 x \ \sqrt{-g} (R - 2\Lambda - \mathscr{L}(\phi))
\end{equation}

where $\phi$ is a general kind of field and $\mathscr{L}(\phi)$ is a general (but known) function of it. We discuss the case of a single field
just for simplicity, but things can be easily generalized even if $\mathscr{L}$ were function of more different fields. Beato-Garcia action is
just a special case of action (\ref{action_general}). It generates the following equations of motion:

\begin{equation}
\label{eq_motion_1_general}
G^\nu_\mu + \Lambda \delta^\nu_\mu = g^{\alpha\nu} \frac{\partial \mathscr{L}}{\partial g^{\mu\alpha}} - \frac{1}{2} \mathscr{L}\delta^\nu_\mu 
\end{equation}

\begin{equation}
\label{eq_motion_2_general}
\nabla_\mu \frac{\partial \mathscr{L}}{\partial (\nabla_\mu \phi)} - \frac{\partial \mathscr{L}}{\partial \phi} = 0
\end{equation}

Assume the second equation to be decoupled from the first: i.e. it does not involve $f(r)$ nor its derivatives (or, equivalently, assume $\tilde{\mathscr{L}}(r)$ to be known. In this case, we can solve it separately, finding $\phi = \phi(r)$ as a known function. Then, plugging this result into eq. (\ref{eq_motion_1_general}), we get

\begin{equation}
\label{solution_general}
f(r) = a - \frac{b}{r} - \frac{\Lambda}{3} r^2 - \frac{1}{r} \int dr \int dr \ r \left(r \phi' \frac{\partial \mathscr{L}}{\partial \phi} + \mathscr{L} \right)  
\end{equation}

where $a$ and $b$ are suitable integration constants. The double integral in this equation is a known function, since we know both
$\mathscr{L}(\phi)$ and $\phi(r)$. The conditions needed to get a SdS solution are

\begin{equation}
\label{cond_1}
a=k
\end{equation}

\begin{equation}
\label{cond_2}
m = \frac{b}{2} + \lim_{r \rightarrow \infty} \frac{1}{2r} \int dr \int dr \ r \left(r \phi' \frac{\partial \mathscr{L}}{\partial \phi} + \mathscr{L} \right)
\end{equation}

Here is the point: $a$ is determined uniquely by the asymptotic topology, while $b$ cannot be determined, since $\mathscr{L}$ contains at least another free parameter. On the other hand, in order to get a regular solution, we must require $b=0$: indeed, $r \rightarrow 0$, the solution goes as

\begin{equation}
\label{f_near_0}
f(r \rightarrow 0) = k - \frac{b}{r} - \frac{1}{6} \left( (\mathscr{L} \circ \phi)(0) + 2 \Lambda \right) r^2 + o(r^3)
\end{equation}

and this satisfies Sakharov criterion only for $b=0$. Hayward \cite{Hayward_1}, Bardeen \cite{Bardeen}, Ansoldi-Nicolini-Smailagic-Spallucci \cite{ANSS} and again Dymnikova \cite{Dymnikova_1}, all of them satisfy this condition. Unfortunately, here we cannot fix $b$, since eq. (\ref{cond_2}) is a single condition which should fix two parameters. Since this cannot be, we must conclude that the most general solution produced by the BGD approach (or any other similar) is still singular.

(By the way, the fact that BGD approach actually contains some "hidden" signularities has already been pointed out in \cite{NBS}, though their focus was different than ours.)

Finally, note that also the result of \cite{Sert}, though with an action written differently from eq. (\ref{action_general}), is affected by the same problem: indeed, they were able (under some assumptions) to write $Y(R)$ as a function of the $r$: $(Y \circ R)(r)$; this makes their Lagrangian in the form $R+\Phi(r)$, with $\Phi(r)$ known, so that the result of the present section is still applicable. On the other hand, \cite{BLZ} and \cite{Elizalde} are able to produce regular black holes (reducing to SdSRN at infinity and satisfying Sakharov criterion), whose Lagrangians are known. These are great results, but should be noted that 1) in both cases the coupling is not minimal and 2) the nonlinear equations of the system cannot be decoupled, so the argument of the present section doesn't hold.

\section{Conclusion}

We discussed the NED paradigm and showed how to generate the most general topological solution. We discussed its general features,
and showed its thermodynamical properties: the First Law of Thermodynamics for black holes arises quite naturally and assumes
the expected form $dU=TdS$. Since this is a general feature of a NED solution and not the result of some specific case (which could be just
a mere coincidence), we interpret it as a clue that NED is a self-consistent picture, which is able to produce some regular solutions.

Unfortunately, BGD procedure is plagued by two main problems: first, the solution generated by a given Lagrangian is unique only as long as we do not introduce cosmological effect (i.e. until $\Lambda=0$); second, as shown in sec. 6, NED models do not avoid, in the most general case, the central singularity\footnote{The result, as shown in sec. 6, is actually more general.}.
These two problems appear to be indications that NED models are not the best tool to generate regular black holes: first, because one expects the relation between the solution and its generating Lagrangian to be unique;  second, because one expects regularity to emerge naturally, not to be imposed by a fine tuning (in our case, $c=0$). Moreover, keeping together our results of sec. 6 and the procedure of \cite{BLZ}, this is a hint that, in order to find regular solutions, one should attack directly the gravitational sector.

Although this objection, however, NED solutions still deserve some interest, since they are actually able to produce (some) regular solutions (and these are unique, if the space is asymptotically Minkowskian). When coming to them, we found a procedure, which greatly simplifies their discussion. Indeed, using the BGD scheme, one is able to find easily only the Hamiltonian $\mathscr{H}(P)$, while one is interested in the Lagrangian $\mathscr{L}(I)$. This requires solving the differential equation

\begin{equation}
\label{diff_sol}
\partial_I \mathscr{L}(I) = \frac{1}{\mathscr{H} \left( I(\mathscr{L}(I) \right)^2}
\end{equation}

which, unless very special cases, is rather complicated. In sec. 5 we showed how this scheme is much simplified, reducing to solve the (typically) algebraic equation (\ref{I(r)}) instead of the differential equation (\ref{diff_sol}).

\bigskip

\bf Acknowledgments. \rm We would like to thank prof. A. Ghosh, prof. L. Vanzo and prof. S. Zerbini for the very profitable conversations.

\section{Appendices}
\appendix
\section{Uniqueness of the Lagrangian $\iff$ uniqueness of the Hamiltonian}

We start from the definitions of $\mathscr{H}$ and $P$ and we operate the variation of $\mathscr{H}(P)$, using both its definitions: we respectively get

\begin{equation}
\label{delta_H_1}
\delta \mathscr{H}(P) = \left( \partial_P \mathscr{H}(P) \right) \delta P
\end{equation}

and

\begin{equation}
\label{delta_H_2}
\begin{split}
\delta \mathscr{H}(P) & = \delta \left( \mathscr{H} \circ P \right) (I) \\
& = \delta \left( 2I (\partial_I \mathscr{L}(I)) - \mathscr{L}(I) \right) \\
& = \frac{1}{\partial_I \mathscr{L}(I)} \delta \left( I (\partial_I \mathscr{L}(I))^2 \right) \\
& = \frac{1}{\partial_I \mathscr{L}(I)} \delta P
\end{split}
\end{equation}

Combining eq. (\ref{delta_H_1}) and eq. (\ref{delta_H_2}) it is then clear that

\begin{equation}
\label{H_L}
\partial_P \mathscr{H}(P) = \frac{1}{\partial_I \mathscr{L}(I)}
\end{equation}

Eq. (\ref{H_L}) implies that, asking $\mathscr{L}(I)$ to be fixed, then $\mathscr{H}(P)$ is fixed too and vice versa (recall that $P$ and $I$ are \it not \rm independent).

\section{ADM energy and ADM quasi local energy function}

We recall that the general expression for the ADM energy is \cite{MTW}

\begin{equation}
\label{C1}
E_{ADM} = \frac{1}{8\pi} \int \epsilon_{\mu\nu\rho\sigma} \nabla^{\rho} \xi^{\sigma} \ dA^{\mu\nu}
\end{equation}

where $A^{\mu\nu}$ is a surface of constant and infinite radius and $\xi^{\sigma}$ is a Killing field. Using this expression for a metric of the form (\ref{metric}) (we already drop out time dependence, since we know that BGD approach only produces static solutions) one has \cite{MTW} \cite{Brewin}

\begin{equation}
\label{C2}
\begin{split}
E_{ADM} &= \frac{1}{2} r^2 f'(r) \mid_{r=\infty} \\
&= E(r) \mid_{r=\infty}
\end{split}
\end{equation}

Where $E(r) \equiv \frac{1}{2} r^2 f'(r)$ is our ADM quasi local energy function. It is easy to see that, for a Schwarzschild-like metric (as we defined it in sec. 2, with $\Lambda=0$) the ADM energy is $E_{ADM}=m$. When the metric is exactly KS, $E(r)=m$ at any $r$.


\begin{thebibliography}{9}

\bibitem{Born_Infeld} M. Born, L. Infeld, \it Foundations of the new field theory\rm, Proc. Roy. Soc. 144, 852 (1934).

\bibitem{Bardeen} J.M. Bardeen, in \it Conference Proceedings of GR5 \rm (Tbilisi, URSS, 1968), p. 174.

\bibitem{Bronnikov} K.A. Bronnikov, \it Regular magnetic black holes and monopoles from nonlinear electrodynamics\rm,
Phys. Rew. {\bf D 63}, 044005 (2001); doi: 10.1103/PhysRevD.63.044005 [gr-qc/0006014].

\bibitem{Elizalde} E. Elizalde, S.R. Hildebrandt \it Family of regular interiors for nonrotating black holes with $T_0^0 = T_1^1$\rm,
Phys. Rev. {\bf D 65}, 124024 (2002); doi: 10.1103/PhysRevD.65.124024 [gr-qc/0202102v2].

\bibitem{Dymnikova} I. Dymnikova, \it Regular electrically charged vacuum structures with de Sitter centre in nonlinear
electrodynamics coupled to general relativity\rm, Class. Quantum Grav. {\bf 21}, 4417 (2004); doi: 10.1088/0264-9381/21/18/009 [gr-qc/0407072].

\bibitem{Hayward_1} S. Hayward, \it Formation and evaporation of non singular black holes\rm,
Phys. Rev. Lett. {\bf 96}, 031103 (2006); doi: 10.1103/PhysRevLett.96.031103 [gr-qc/0506126].

\bibitem{NSS} P. Nicolini, A. Smailagic, E. Spallucci, \it Noncommutative geometry inspired Schwarzschild black hole\rm, Phys. Lett. {\bf B 632}, 547 (2006); doi: 10.1016/j.physletb.2005.11.004 [gr-qc/0510512].

\bibitem{ANSS} S. Ansoldi, P. Nicolini, A. Smailagic, E. Spallucci, \it Noncommutative geometry inspired charged black holes\rm, Phys. Lett. {\bf B 645}, 261 (2007); doi: 10.1016/j.physletb.2006.12.020 [gr-qc/0612035v1].

\bibitem{Modesto} S. Hossenfelder, L. Modesto, I. Prémont-Schwarz, \it A model for non-singular black hole collapse and evaporation\rm, Phys. Rev. {\bf D 81}, 044036 (2010); doi: 10.1103/PhysRevD.81.044036 [gr-qc/0912.1823v3].

\bibitem{CMSZ} G. Cognola, R. Myrzakulov, L. Sebastiani, S. Zerbini, \it Einstein gravity with Gauss-Bonnet entropic corrections\rm, Phys. Rew. {\bf D 88}, 024006 (2013); doi: 10.1103/PhysRevD.88.024006 [gr-qc/1304.1878v2].

\bibitem{Dymnikova_2} I. Dymnikova, E. Galaktionov, \it Regular rotating electrically charged black holes and solitons in nonlinear electrodynamics minimally coupled to gravity\rm, Class. Quantum Grav. {\bf 32}, 165015 (2015); doi: 10.1088/0264-9381/32/16/165015 [gr-qc/1510.01353v1].

\bibitem{Culetu} H. Culetu, \it Microscopic corrections to Schwarzschild spacetime \rm (2015); [gr-qc/1508.07570v2].

\bibitem{Maeda} G. Kunstatter, H. Maeda, T. Taves, \it New two-dimensional effective actions for non-singular black holes \rm (2015) [gr-qc/1509.06746v2].

\bibitem{Horava} P. Horava, \it Membranes at quantum criticality\rm, JHEP {\bf 0903} 20 (2009); doi: 10.1088/1126-6708/2009/03/020 [hep-th/0812.4287v3].

\bibitem{Horava_1} P. Horava, \it Quantum gravity at a Lifshitz point\rm, Phys. Rev. {\bf D 79} 084008 (2009); doi: 10.1103/PhysRevD.79.084008 [hep-th/0901.3775v2].

\bibitem{KS} A. Kehagias, K. Sfetsos, \it The black hole and FRW geometries of non-relativistic gravity\rm, Phys. Lett. {\bf B 678}, 123 (2009); doi: 10.1016/j.physletb.2009.06.019 [hep-th/0905.0477v1].

\bibitem{CCO} R.G. Cai, L.M. Cao, N. Ohta, \it Black holes in gravity with conformal anomaly and logarithmic term in black hole entropy\rm, JHEP {\bf 1004} 082 (2010); doi: 10.1007/JHEP04(2010)082 [hep-th/0911.4379v2].

\bibitem{Pradhan} P. Pradhan, \it Area (or entropy) product formula for a regular black hole \rm (2015); [gr-qc/1512.06187].

\bibitem{Ma} M.S. Ma, \it Magnetically charged regular black hole in a model of nonlinear electrodynamics\rm, Annals Phys. {\bf 362} 529 (2015); doi:  10.1016/j.aop.2015.08.028 [gr-qc/1509.05580].

\bibitem{Johannsen} T. Johannsen, \it Regular black hole metric with three constants of motion\rm, Phys. Rev. {\bf D 88} 044002 (2013); doi: 10.1103/PhysRevD.88.044002 [gr-qc/1501.02809v2].

\bibitem{Rodrigues} M.E. Rodrigues, J.C. Fabris, E.L.B. Junior, G.T. Marques \it Generalization of regular black holes in General Relativity to $f(R)$ gravity \rm; [gr-qc/1601.00471].

\bibitem{Beato} E. Ayon–Beato, A. Garcia, \it Regular black hole in general relativity coupled to nonlinear electrodynamics\rm,
Phys. Rew. Lett. {\bf 80}, 5056 (1998); doi: 10.1103/PhysRevLett.80.5056 [gr-qc/9911046v1].

\bibitem{Beato_1} E. Ayon-Beato, A. Garcia, \it The Bardeen model as a non linear magnetic monopole\rm,
Phys. Lett. {\bf B 493}, 149 (2000); doi: 10.1016/S0370-2693(00)01125-4 [gr-qc/0009077].

\bibitem{Dereli} T. Dereli, Ö. Sert, \it Non-minimal $ln(R) F^2$ couplings of electromagnetic fields to gravity: static, spherically symmetric solutions\rm, Eur. Phys. J. {\bf C 71}, 1589 (2011); doi: 10.1140/epjc/s10052-011-1589-2 [gr-qc/1102.3863v1].

\bibitem{Sert} Ö. Sert, \it Regular black hole solutions of the non-minimally coupled $Y(R)F^2$ gravity \rm (2015); [gr-qc/1512.01172v1].

\bibitem{Balakin} A.B. Balakin, J.P.S. Lemos, \it Non-minimal coupling for the gravitational and electromagnetic fields: a general system of equations\rm, Class. Quantum Grav. {\bf 22}, 1867 (2005); doi: 10.1088/0264-9381/22/9/024 [gr-qc/0503076v2].

\bibitem{BLZ} A.B. Balakin, J.P.S. Lemos, A.E. Zayats, \it Regular nonminimal magnetic black holes in spacetimes with a cosmological constant \rm (2015); [gr-qc/1512.02653v1].

\bibitem{Zayats} A.B. Balakin, A.E. Zayats, \it Non-minimal Wu-Yang monopoles\rm, Phys. Lett. {\bf B 644}, 294 (2006); doi: 10.1016/j.physletb.2006.12.002 [gr-qc/0612019].

\bibitem{Horndeski} G.W. Horndeski, \it Static spherically symmetric solutions to a system of generalized Einstein-Maxwell field equations\rm, Phys. Rev. {bf D 17}, 391 (1978); doi: http://dx.doi.org/10.1103/PhysRevD.17.391.

\bibitem{Drummond} I.T. Drummond, S.J. Hathrell, \it QED vacuum polarization in a background gravitational field and its effect on the velocity of photons\rm, Phys. Rew. {\bf D 22}, 343 (1980); doi: 10.1103/PhysRevD.22.343.

\bibitem{BLP} D.R. Brill, J. Louko, P. Peldàn, \it Thermodynamics of (3+1)-dimensional black holes with toroidal or higher genus
horizon\rm, Phys. Rev. {\bf D 56}, 3600 (1997); doi: 10.1103/PhysRevD.56.3600 [gr-qc/9705012].

\bibitem{Mann} R.B. Mann, \it Pair production of topological anti-de Sitter black holes\rm, Class. Quantum Grav. {\bf 14}, L109 (1997);
doi: 10.1088/0264-9381/14/5/007 [gr-qc/9607071].

\bibitem{Vanzo} L. Vanzo, \it Black holes with unusual topology\rm, Phys. Rew. {\bf D 56}, 6475 (1997); doi:10.1103/PhysRevD.56.6475 [gr-qc/9705004]

\bibitem{Hawking_Ellis} S.W. Hawking, G.F.R. Ellis, \it The large scale structure of spacetime\rm, Cambridge University Press (1973).

\bibitem{Gibbons} G.W. Gibbons, D.L. Wiltshire, \it Black holes in Kaluza-Klein theory\rm, Ann. Phys. {\bf 167}, 1 (1986); doi: 10.1016/S0003-4916(86)80012-4.

\bibitem{Park} Y.S. Myung, Y.W. Kim, Y.J. Park, \it Thermodynamics of regular black hole\rm, Gen. Rel. Grav {\bf 41}, 1051 (2009);
doi: 10.1007/s10714-008-0690-9 [gr-qc/0708.3145].

\bibitem{Akbar} M. Akbar, N. Salem, S.A. Hussein, \it Thermodynamics of the Bardeen regular black hole\rm, Chin. Phys. Lett.
{\bf 29}, 7, 070401 (2012); doi: 10.1088/0256-307X/29/7/070401.

\bibitem{Saneesh} S. Saneesh, V.C. Kuriakose, \it Spectroscopy and thermodynamics of a regular black hole\rm, Int. J. Phys {\bf 54},
3162 (2015); doi: 10.1007/s10773-015-2555-9.

\bibitem{Hawking} S.W. Hawking, \it Particle creation by black holes\rm, Comm. Math. Phys. {\bf 43}, 199 (1975); doi: 10.1007/BF02345020.

\bibitem{Hayward} S. Hayward, \it General laws of black hole dynamics\rm,
Phys. Rew. {\bf D 49}, 6467 (1994); doi: 10.1103/PhysRevD.49.6467 [gr-qc/9303006].

\bibitem{Sakharov} A. Sakharov, \it Initial stage of an expanding universe and appearance of a nonuniform distribution of matter\rm,
Sov. Phys. JETP {\bf 22}, 241 (1966).

\bibitem{Dymnikova_1} I. Dymnikova, \it Spherically symmetric space time with the regular de Sitter center\rm,
Int. J. Mod. Phys. {\bf 1015} (2003); doi: 10.1142/S021827180300358X.

\bibitem{Rinaldi} M. Rinaldi, \it Black holes with non-minimal derivative coupling\rm, Phys. Rew. {\bf D 86}, 084048 (2012); doi: 10.1103/PhysRevD.86.084048 [gr-qc/1208.0103v5].

\bibitem{NBS} M. Novello, S.E.P. Bergliaffa, J.M. Salim, \it Singularities in general relativity coupled to nonlinear electrodynamics\rm, Class. Quantum Grav {\bf 17}, 18 (2000); doi: 10.1088/0264-9381/17/18/316 [gr-qc/0003052].

\bibitem{MTW} C.W. Misner, K.S. Thorne, J.A. Wheeler, \it Gravitation\rm, W.H. Freeman \& Co. (1973).

\bibitem{Brewin} L. Brewin, \it A simple expression for the ADM mass\rm, Gen. Rel. Grav. {\bf 39}, 521 (2007); doi: 10.1007/s10714-007-0403-9 [gr-qc/0609079].

\end{thebibliography}
\end{document}